\documentclass[runningheads]{llncs}
\usepackage[T1]{fontenc}
\usepackage{graphicx}
\usepackage[misc]{ifsym}
\usepackage{adjustbox}
\usepackage{amsmath}
\usepackage{amssymb}
\usepackage{upgreek}
\usepackage{multirow}
\usepackage{booktabs}  
\usepackage[font={footnotesize}]{caption}
\usepackage[table,xcdraw]{xcolor}

\newcommand{\squeezeup}{\vspace{-3mm}}

\begin{document}
%
%\title{Integrating Laplace’s equation within a deep
%learning framework to model the sheet-like geometry of the cortex in ex vivo MRI}
%\title{Integrating Laplace’s equation within a deep
%learning framework to encourage segmentation of deep sulci in cortical gray matter}
%

\title{Improved Segmentation of Deep Sulci in Cortical Gray Matter Using a Deep Learning Framework Incorporating Laplace's Equation}

\titlerunning{Integrating Laplacian Constraints within a Cortical Segmentation Network}
\author{Sadhana Ravikumar\inst{1}\textsuperscript{(\Letter)} \and
Ranjit Ittyerah\inst{1} \and
Sydney Lim\inst{1} \and
Long Xie\inst{1} \and
 Sandhitsu Das\inst{1} \and
 Pulkit Khandelwal\inst{1} \and
 Laura E.M. Wisse\inst{2} \and
 Madigan L. Bedard \inst{1} \and
 John L. Robinson\inst{1} \and
 Terry Schuck\inst{1} \and
 Murray Grossman\inst{1} \and
 John Q. Trojanowski \inst{1} \and
 Edward B. Lee\inst{1} \and
 M. Dylan Tisdall\inst{1} \and
 Karthik Prabhakaran\inst{1} \and
 John A. Detre\inst{1} \and
 David J. Irwin\inst{1} \and
 Winifred Trotman\inst{1} \and
 Gabor Mizsei\inst{1} \and
 Emilio Artacho-P\'erula\inst{3} \and
 Maria Mercedes Iñiguez de Onzono Martin\inst{3} \and
 Maria del Mar Arroyo Jim\'enez\inst{3} \and
 Monica Mu\~noz\inst{3} \and
 Francisco Javier Molina Romero\inst{3} \and
 Maria del Pilar Marcos Rabal\inst{3} \and
 Sandra Cebada-S\'anchez\inst{3} \and
 Jos\'e Carlos Delgado Gonz\'alez\inst{3} \and
 Carlos de la Rosa-Prieto\inst{3} \and
 Marta C\'orcoles Parada \inst{3} \and
 David A. Wolk\inst{1} \and
 Ricardo Insausti\inst{3} \and
 Paul A. Yushkevich\inst{1}\textsuperscript{(\Letter)} }
\authorrunning{S Ravikumar et al.}
%% % First names are abbreviated in the running head.
%% % If there are more than two authors, 'et al.' is used.
%% %
\institute{University of Pennsylvania, USA \\
\email{\{ravikums,pauly2\}@pennmedicine.upenn.edu} \and
 Lund University, Sweden \and
 University of Castilla-La Mancha, Spain}
% \email{\{abc,lncs}@uni-heidelberg.de}}
%
\maketitle              % typeset the header of the contribution
\begin{abstract}
When developing tools for automated cortical segmentation, the ability to produce topologically correct segmentations is important in order to compute geometrically valid morphometry measures. In practice, accurate cortical segmentation is challenged by image artifacts and the highly convoluted anatomy of the cortex itself. To address this, we propose a novel deep learning-based cortical segmentation method in which prior knowledge about the geometry of the cortex is incorporated into the network during the training process. We design a loss function which uses the theory of Laplace's equation applied to the cortex to locally penalize unresolved boundaries between tightly folded sulci. Using an \textit{ex vivo} MRI dataset of human medial temporal lobe specimens, we demonstrate that our approach outperforms baseline segmentation networks, both quantitatively and qualitatively.

\end{abstract}

\section{Introduction}
\squeezeup
Segmentation of the cerebral cortex from MRI is an important first step in many neuroimaging pipelines such as quantitative morphometry analyses aimed at understanding the pathophysiology of neurological disorders. Automated segmentation methods applied to the cortex are challenged by various artifacts such as image noise, partial volume effects and intensity inhomogeneities which make accurate identification of the tissue boundaries difficult and result in geometrically inaccurate cortical reconstructions. The cerebral cortex or gray matter (GM) can be defined as the space between two cortical surfaces; the pial surface which separates the GM from the surrounding cerebrospinal fluid (CSF), and the white matter (WM) surface which separates the GM from WM. The cortex has a complex geometry and is often modelled as a highly folded 2D sheet, with spatially varying curvature and thickness \cite{Osechinskiy2012}. Geometrically accurate segmentation of the cortex requires accurate reconstruction of both the WM and pial cortical surfaces, complete with all cortical folds and narrow sulci. A commonly used simplification when solving cortical surface reconstruction problems is to view the cortical surfaces as having the topology of a 3D sphere \cite{fischl2012freesurfer}. However, unless explicitly corrected for, imaging artifacts often introduce topological defects in the resulting surface reconstructions. Defects due to partial volume effects are particularly apparent in tightly folded sulci and result in opposing banks of sulci appearing fused together. This creates either `bridged' or `unresolved' sulci in the resulting cortical reconstruction, which cause errors in downstream quantitative brain morphometry measures such as cortical thickness. While topological defects can be corrected by manual editing, these checks can be time-consuming. 

Topology-corrected reconstruction of cortical surfaces is a well studied topic in \textit{in vivo} neuroimaging literature, and several state-of-the-art methods have been developed to address this problem \cite{fischl2012freesurfer,kim2005,Han2004}. The widely used Freesurfer framework employs a mesh-based approach to topology-correction which consists of two main steps \cite{fischl2012freesurfer}. First, the inner WM surface is generated by applying mesh tessellation to a volumetric WM segmentation that has been corrected for topological defects \cite{fischl2012freesurfer}. Second, this WM surface is expanded using a deformable surface model to reconstruct the outer, pial surface, while ensuring that the topology of the initial surface is preserved \cite{fischl2012freesurfer,Han2004,kim2005}. In recent years, elements of the FreeSurfer pipeline have been implemented as deep learning networks, resulting in significant speedups \cite{ma2021pialnn,henschel2020fastsurfer,cruz2021deepcsr,hoopes2021topofit}. However, these frameworks still either require the time-consuming post-processing step of topology correction \cite{cruz2021deepcsr,henschel2020fastsurfer} or rely on a predefined initial mesh with the correct, spherical topology to reconstruct the cortical surface \cite{hoopes2021topofit,ma2021pialnn}. 

In this work, we are specifically interested in developing an automated cortical segmentation method that can be applied to \textit{ex vivo} brain MRI datasets to generate geometrically valid models of the cortex. Instead of a mesh deformation-based approach, we propose a novel volumetric deep image segmentation method that learns to segment the cortex while explicitly modeling the `sheet-like' geometry of the cortex. In \textit{ex vivo} studies, it is common to image only a portion of the brain hemisphere, thus violating the assumption of spherical WM topology made by mesh-based approaches. Furthermore, many of the \textit{in vivo} cortical segmentation methods contain algorithms that are optimized for data with a standard 1 mm voxel size \cite{fischl2012freesurfer}, and would result in unrealistic computational times if applied to high-resolution \textit{ex vivo} MRI datasets. As a result, existing methods are not easily applicable to \textit{ex vivo} MRI scans.  As far as we know, no prior work has focused on topology correction of \textit{ex vivo} cortical segmentations. 

Previous studies have used the concept of Laplace's equation as a tool for modelling the cortex \cite{jones2000three,Osechinskiy2012,kim2005}. By setting different boundary conditions at the WM/GM and GM/CSF interfaces, Laplace's equation can be solved within the GM volume to generate a laminar `potential' field that smoothly varies in value depending on its distance between the two cortical surfaces. The gradient of the Laplacian field can be used to compute cortical thickness \cite{jones2000three} and defines an expansion path which guarantees topology-preserving deformation between the WM and pial cortical surfaces \cite{kim2005,Osechinskiy2012}. Building on this idea and the success of deep convolutional neural networks (CNN) in medical image segmentation tasks, here we design a differentiable numerical solver for Laplace's equation and incorporate it within a deep segmentation framework to locally impose a Laplacian mapping between the predicted WM and pial surfaces. We train the segmentation network by comparing the predicted tissue segmentations and corresponding Laplacian field maps with the equivalent ground truth images, thus penalizing self-intersections in the predicted segmentations.  Our results show that when compared to a baseline network trained without Laplacian constraints, our method is able to better reconstruct the intrinsic, layered geometry of the cortex. To our knowledge, this is the first time that an iterative numerical solver has been incorporated within a cortical segmentation network to directly compute Laplacian fields in an end-to-end setting.
 
\section{Methods}
\begin{figure}[ht]
    \centering
    \includegraphics[width =0.9\textwidth]{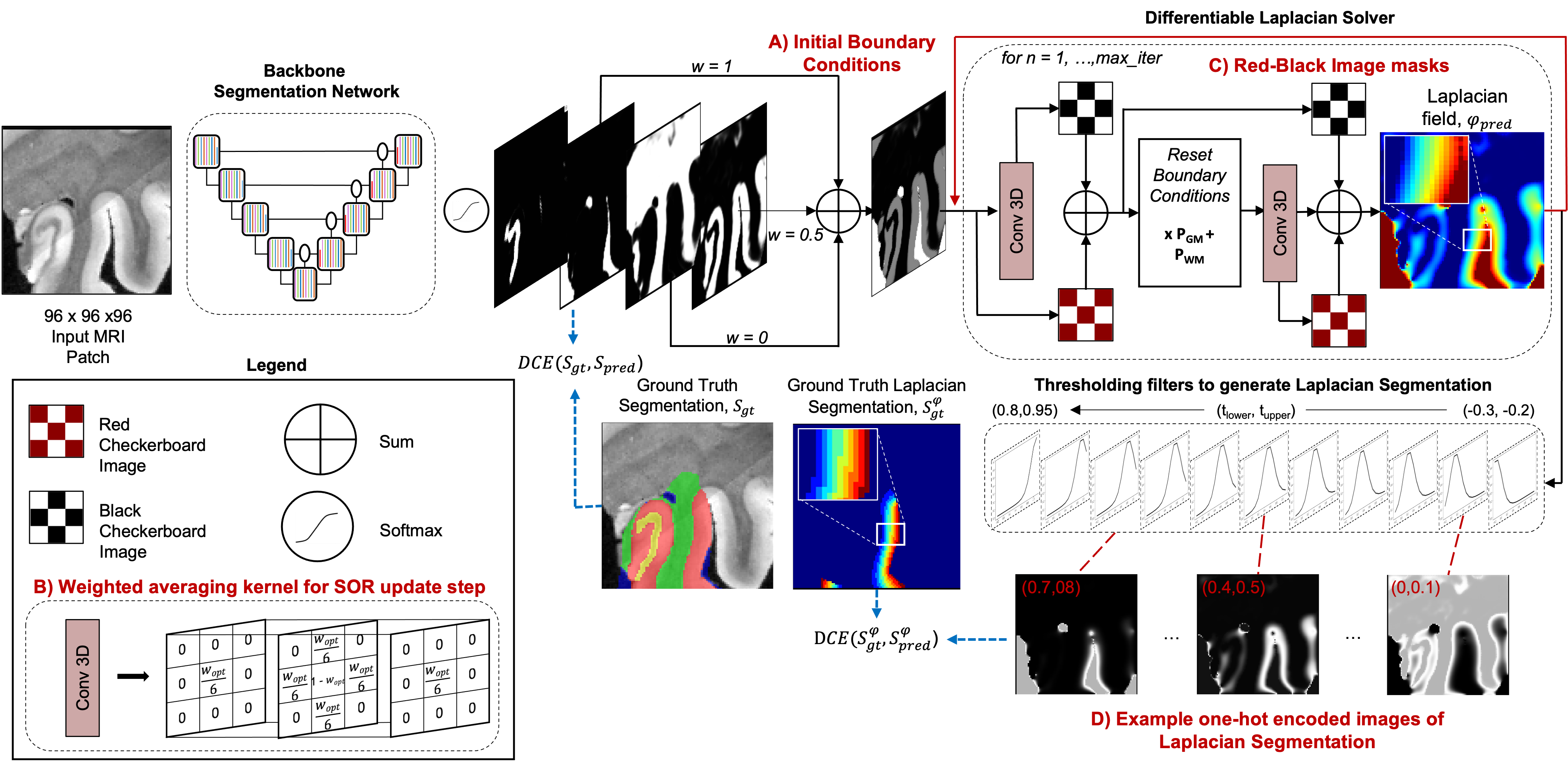}
    \setlength{\belowcaptionskip}{-15pt}
    \caption[Network Architecture]{Schematic illustration of the proposed framework. A differentiable numerical solver for Laplace's equation, based on the successive over relaxation (SOR) algorithm, is incorporated within end-to-end training of a segmentation network.}
    \label{fig:nnUNET_SOR}
\end{figure}

\squeezeup
As illustrated in Fig. \ref{fig:nnUNET_SOR}, our proposed framework builds upon any given backbone segmentation network (Sect. \ref{backbone}) by appending a numerical solver for Laplace's equation to the output of the network. We reformulate the solver to be differentiable with respect to the input image to allow for gradient-based learning, used within standard CNN training (Sect. \ref{lsolver}). In addition to the standard tissue segmentation loss, we design a loss function which compares the predicted Laplacian field to the solution of Laplace's equation applied to the ground truth cortical segmentation, which is assumed to have correct topology (Sect. \ref{lossfunc}).

\subsection{Backbone Segmentation Network}
\label{backbone}
The proposed Laplacian solver is compatible with any semantic segmentation network since it only relies on the segmentation map output by the backbone network. We conducted experiments using two backbone networks for 3D image segmentation: the state-of-art nnU-Net framework \cite{isensee2021nnu} based on the U-net architecture, and nnFormer \cite{zhou2021nnformer}, a variant framework based on the recently popular transformer architecture. Both frameworks use image patches, deep supervision, and a variety of data-augmentation techniques to train the network \cite{isensee2021nnu}.

\subsection{Differentiable Laplacian Solver}
\label{lsolver}
To compute the Laplacian field corresponding to the cortical segmentation predicted based on a given input patch, the iterative solver for Laplace's equation is appended after the final layer of the backbone network. Laplace's equation, $\Delta\varphi = 0$, is a second-order partial differential equation, where $\Delta$ is the Laplacian operator ($\partial^2_{xx} + \partial^2_{yy} + \partial^2_{zz}$) and $\varphi$ is a twice-differentiable, real-valued function. To solve Laplace's equation within a domain (in our case, the GM volume), specific conditions need to be set that the Laplacian field $\varphi$ must satisfy at the boundaries of the domain. In our case, we set $\varphi_{x,y,z} = 0$ at the GM/WM boundary and $\varphi_{x,y,z} = 1$ at the GM/pial boundary. Voxels within the GM domain are initialized with $\varphi_{x,y,z} = 0.5$. 
 
Given these boundary conditions, the Laplacian field can be approximated using the finite-difference method, which is solved using an iterative numerical solver. Here, we use the Successive Over Relaxation (SOR) algorithm, a variant of the Gauss-Seidel method, to solve for the Laplacian field \cite{hansen1992numerical}. In the Gauss-Seidel method, given initial values for all the voxels in an image, at each iteration, the new value for a particular voxel within the GM volume, $ \tilde{\varphi}_{xyz}$, is computed by taking the weighted sum of the most recently updated values of its six neighboring voxels (Eq. \ref{eq:GS}). The superscript $n$ refers to the iteration of the algorithm and the subscripts are the voxel indices.
%$ \tilde{\varphi}_{xyz}^{n} = \frac{1}{6}(\varphi^{n}_{x-1,y,z} + \varphi^{n-1}_{x+1,y,z} +  \varphi^{n}_{x,y-1,z} + \varphi^{n-1}_{x,y+1,z} + \varphi^{n}_{x,y,z-1} + \varphi^{n-1}_{x,y,z+1})$
\begin{equation}
\label{eq:GS}
    \tilde{\varphi}_{xyz}^{n} = \frac{1}{6}(\varphi^{n}_{x-1,y,z} + \varphi^{n-1}_{x+1,y,z} + \\ \varphi^{n}_{x,y-1,z} + \varphi^{n-1}_{x,y+1,z} + \varphi^{n}_{x,y,z-1} + \varphi^{n-1}_{x,y,z+1})
\end{equation}

The SOR algorithm accelerates this approach by taking, at each iteration $n$, the weighted sum of the current solution and the solution from the previous iteration (Eq. \ref{SOReq}). The over-relaxation parameter, $\omega$ accelerates the rate of convergence of the Gauss-Seidel method when $1 < \omega < 2$ \cite{yang2007optimal}. It has been shown that the optimum value, $\omega_{opt}$ is given by $\omega_{opt} = \frac{2}{1 + sin(\frac{\pi}{N + 1})}$ where $N$ is the minimum dimension of the input grid \cite{yang2007optimal}.

\begin{equation}
\label{SOReq}
\begin{split}
    \varphi^{n}_{x,y,z} = (1 - \omega_{opt})\varphi^{n-1}_{x,y,z} + \frac{\omega_{opt}}{6}(\varphi^{n}_{x-1,y,z} + \varphi^{n-1}_{x+1,y,z} + \\ \varphi^{n}_{x,y-1,z} + \varphi^{n-1}_{x,y+1,z} + \varphi^{n}_{x,y,z-1} + \varphi^{n-1}_{x,y,z+1})
    \end{split}
\end{equation}

Instead of updating the value of each voxel in the image serially, computation of Eq. \ref{SOReq} can be parallelized using the Red-Black SOR approach, wherein the voxels in an image are divided into `red' and `black' following a checkerboard pattern \cite{epicoco2012performance}. During the update step, the `red' voxels only depend on the values of the `black' voxels and vice versa. Therefore, at each iteration, the Laplacian solution is updated in two steps; first, the update equation is applied to all of the black voxels in parallel and second, the update equation is applied to all of the red voxels in parallel, using the updated values computed at the black voxels. After convergence, the resulting Laplacian field contains values within the GM volume increasing smoothly from 0 at the WM surface to 1 at the pial surface.\\ 

To incorporate this numerical solver within a CNN, an important consideration is that the computations used to generate the Laplacian field must be differentiable with respect to the predicted tissue class probabilities to allow for back-propagation of the final loss through the network. To this end, we initialized the boundary conditions for the Laplacian solver by taking a weighted sum of the GM, WM and background probability maps, with weights of 0.5, 0 and 1 respectively (Fig. \ref{fig:nnUNET_SOR}A). Additionally, the SOR update step (Eq. \ref{SOReq}) was reformulated as a 1$\times$1 convolutional layer with fixed neighborhood weights specified in a 3$\times$3$\times$3 kernel (Fig. \ref{fig:nnUNET_SOR}B). The voxels in the image were divided into a red and black grid by generating 3D `red' ($mod(x + y + z,2) = 0$), and `black' ($mod(x + y + z,2) = 1$) binary checkerboard images which were applied as image masks to retain values of interest after each convolutional layer (Fig. \ref{fig:nnUNET_SOR}C). Lastly, the maximum number of iterations for the Laplacian solver was empirically set to 60, as a trade-off between computational time and convergence of the Laplacian solution. Since the solver numerically computes the solution to Laplace's equation, it does not introduce any additional training parameters within the network.

\subsection{Loss Function}
\label{lossfunc}
To train the model, the backbone networks compare the predicted tissue segmentation, $S_{pred}$ with the ground truth cortical segmentation, $S_{gt}$ using a combination of Dice and cross-entropy loss, $DCE(S_{gt},S_{pred})$ \cite{isensee2021nnu}. We introduce an additional loss term which compares the Laplacian field computed from the predicted tissue segmentation, $\varphi_{pred}$, with the solution of Laplace's equation applied to the ground truth segmentation, $\varphi_{gt}$. To simplify comparison between the predicted Laplacian field and the ground truth solution, we convert the Laplacian field computed by the solver to a multi-label segmentation, $S^{\varphi}_{pred}$ using a series of thresholding functions. The advantage of this approach is that it enables the use of the same combined Dice and cross-entropy loss used by the backbone network on the outputs of the Laplacian solver, instead of the mean square error loss, thereby allowing us to equally weight the two loss terms: $\mathcal{L} = DCE(S_{gt},S_{pred}) + DCE(S^{\varphi}_{gt},S^{\varphi}_{pred})$. To threshold the Laplacian field and create a multi-label segmentation in a differentiable way, the computed Laplacian field is passed through a product of two sigmoid functions, $(1 + e^{-\beta(x-t_{lower})})^{-1} \times (1 + e^{\beta(x-t_{upper})})^{-1}$, which together create a `band-pass' thresholding filter. $\beta$ controls the steepness of the filter, and $t_{lower}$ and $t_{upper}$ control the domain of the filter. Each filter creates an image that has values close to 1 for voxels in the Laplacian field lying within the domain of the filter, and values close to 0 otherwise. Therefore, by varying the lower and upper threshold values, a one-hot encoded image can be created, where each channel corresponds to a different label along the laminar axis of the GM. A multi-label Laplacian segmentation is then generated by applying the argmax operation to the computed one-hot encoded image (Fig. \ref{fig:nnUNET_SOR}D).

\section{Experiments}

\subsection{Dataset}
\label{data}
\paragraph{MRI Image Acquisition} To train and evaluate the proposed framework, we used \textit{ex vivo} images of intact temporal lobe specimens, obtained from 27 brain donors from either the brain bank operated by the National Disease Research Interchange, or autopsies performed at the University of Pennsylvania Center for Neurodegenerative Disease Research (CNDR) and the University of Castilla-La Mancha (UCLM) Human Neuroanatomy Laboratory (HNL) in Spain. Brain specimens were obtained in accordance with the University of Pennsylvania Institutional Review Board guidelines, and the Ethical Committee of UCLM. Where possible, pre-consent during life and, in all cases, next-of-kin consent at death was given. Following 4+ weeks of fixation, the tissue specimens were scanned overnight on a 9.4 Tesla 31 cm bore MRI scanner using a T2-weighted, multi-slice spin echo sequence (TE = 9330ms, TR = 23ms), with a resolution of 0.2 x 0.2 x 0.2 mm$^3$. Following image acquisition, the images were corrected for bias field non-uniformity and normalized to a common intensity range of [0,1000]. In our work, we are specifically interested in segmenting the medial temporal lobe (MTL), a region affected early in Alzheimer's Disease. Therefore, to facilitate semi-automated MTL segmentation, each scan was re-oriented so that the long axis of the MTL aligned with the anterior-poster direction. 

\paragraph{Ground Truth Tissue Segmentation} To generate 3D segmentations of the MTL cortex, we adopted a semi-automatic interpolation technique \cite{Ravikumar2019}. The boundary of the MTL was manually traced approximately every 3 mm (i.e. 12-15 slices per dataset). Given the subset of labeled slices, the interpolation method uses contour and intensity information to compute the intermediate segmentations. This algorithm was applied iteratively, allowing the interpolated result to be reviewed and manually edited at each step to refine the segmentation. When editing, we ensured that in narrow and bridged sulci, the full extent of each sulcus was correctly labeled as background. Additionally, in a small region surrounding the MTL, the white matter and background voxels were semi-automatically labeled using a combination of intensity-based thresholding and morphological operations. The ground truth segmentations also contain a separate label for the stratum radiatum lacunosum moleculare (SRLM), which is the thin, hypo-intense layer within the hippocampus. We note that the ground truth segmentations only cover the region in the image encompassing the MTL, and not the entirety of the \textit{ex vivo} MRI scan. 

\paragraph{Ground Truth Laplacian maps} The proposed framework requires the Laplacian field maps corresponding to the ground truth segmentations to train the model. To solve Laplace's equation within the ground truth GM volume, we used the iterative finite-differences approach, as implemented in \cite{dekraker2018unfolding}. This implementation employs a 26-neighbour average to compute the updated potential field and terminates the numerical solver when the Laplacian field change is below a specified threshold (sum of changes < 0.001\% of total volume). To initialize the solver, source and sink boundary conditions were semi-manually labeled as the WM and pial surfaces of the MTL respectively. We note that the hippocampus voxels were not included in the GM domain of Laplace's equation. 

\subsection{Implementation Details}
\label{sec:Implementation}
We used Pytorch 1.9.1 and Nvidia Quadro RTX 500 GPUs to train the models. We implemented the differentiable Laplacian solver within the standardized training framework presented in \cite{isensee2021nnu} that is employed by both backbone networks. The framework includes pre-processing, automated hyper-parameter selection and fixed techniques for data augmentation. In our experiments, we made a few modifications to the default training parameters. First, since the ground truth segmentations only cover a portion of the input \textit{ex vivo} MRI scans, we set the \textit{ignore\_label} parameter in the loss function to 0 to exclude the unlabeled background voxels from the training process. Additionally, we increased the \textit{oversample\_foreground} parameter such that only foreground patches are sampled during training. Lastly, we used an input patch size of 96 x 96 x 96, to encourage the network to learn more local image features instead of larger contextual information, like the anatomical boundaries of the MTL. The networks were trained with a batch size of 2 (nnFormer) and 4 (nnU-Net), for 250 epochs in a five-fold cross validation setting. We used the results of the first fold to tune the network parameters. Consistent with the evaluation scheme used within nnU-Net, we aggregated the results across the remaining four folds for reporting test accuracy. We tested the performance of the network when using either 5 or 10 class labels (i.e. laminar layers) for generating the Laplacian segmentation. We found that increasing the number of class labels and training the network using Laplacian segmentations with a denser number of laminar layers improved the network's ability to detect obscured sulci. To convert the Laplacian fields to segmentations, we used $\beta = 10$ (softmax scaling parameter) and selected evenly spaced thresholds spanning the [0,1] range of the Laplacian field. More specifically, the following lower ($t_{lower}$) and upper ($t_{upper}$) threshold values were used: [(-0.3,-0.2),(0,0.1),(0.1,0.2),(0.2,0.3),(0.3,0.4),(0.4,0.5),(0.5,0.6),(0.6,0.7),(0.7,0.8),\\(0.8,0.95),(0.95,1.05)]. Additionally, we tested the effect of increasing the weight given to the Laplacian segmentation loss relative to the tissue segmentation loss and found that it had minimal effect on cortical segmentation accuracy.  

\subsection{Evaluation}
We compared the performance of our approach with the performance of the corresponding backbone segmentation networks, trained only with the tissue segmentation loss. We measured segmentation accuracy by computing the DSC between the predicted and ground truth tissue segmentations, and Laplacian field segmentations within the MTL region of interest. We also report the symmetric Hausdorff Distance (HD) 95$^{th}$ percentile between the predicted and ground truth segmentation of the MTL cortex. Since the numerical solvers used to generate the ground truth Laplacian fields and embedded in the network leverage different finite-difference approximation methods, during evaluation, we re-computed the Laplacian field for both the ground truth and predicted cortical segmentations using 120 iterations of the SOR solver used by the network, and computed the corresponding Laplacian segmentation with 5 laminar layers.

In a secondary analysis we evaluated the effect of introducing the Laplacian constraint on downstream cortical thickness measures. We applied the nnU-Net models to a dataset of 36 temporal lobe specimens obtained from individuals not included in the training dataset. For each specimen, we quantified MTL thickness at 6 manually identified landmarks corresponding to the anterior and posterior locations of MTL subregions Brodmann Area (BA) 35, BA36 and the parahippocampal cortex (PHC). We chose these subregions since they typically lie along the banks of the collateral sulcus (CS), and are therefore mostly likely to be affected by topological errors in the segmentation. For each location, we extracted the GM segmentation surrounding the landmark and measured cortical thickness using the pipeline described in \cite{Wisse2021}. In brief, given the GM segmentation surrounding a landmark, a maximally inscribed sphere is computed using Voronoi skelentonization \cite{Ogniewicz1995}, and the diameter of the sphere gives the thickness at that landmark. We compared thickness measurements obtained when using automatic GM segmentations generated by the baseline nnU-Net and the proposed model (\textit{nnU-Net+Laplacian}), and reference thickness measurements computed using semi-automatic segmentations of the GM in terms of Pearson's correlation and the average fixed-raters Intra-Class Correlation Coefficient (ICC). 

\begin{figure}[ht]
    \centering
    \includegraphics[width =\textwidth]{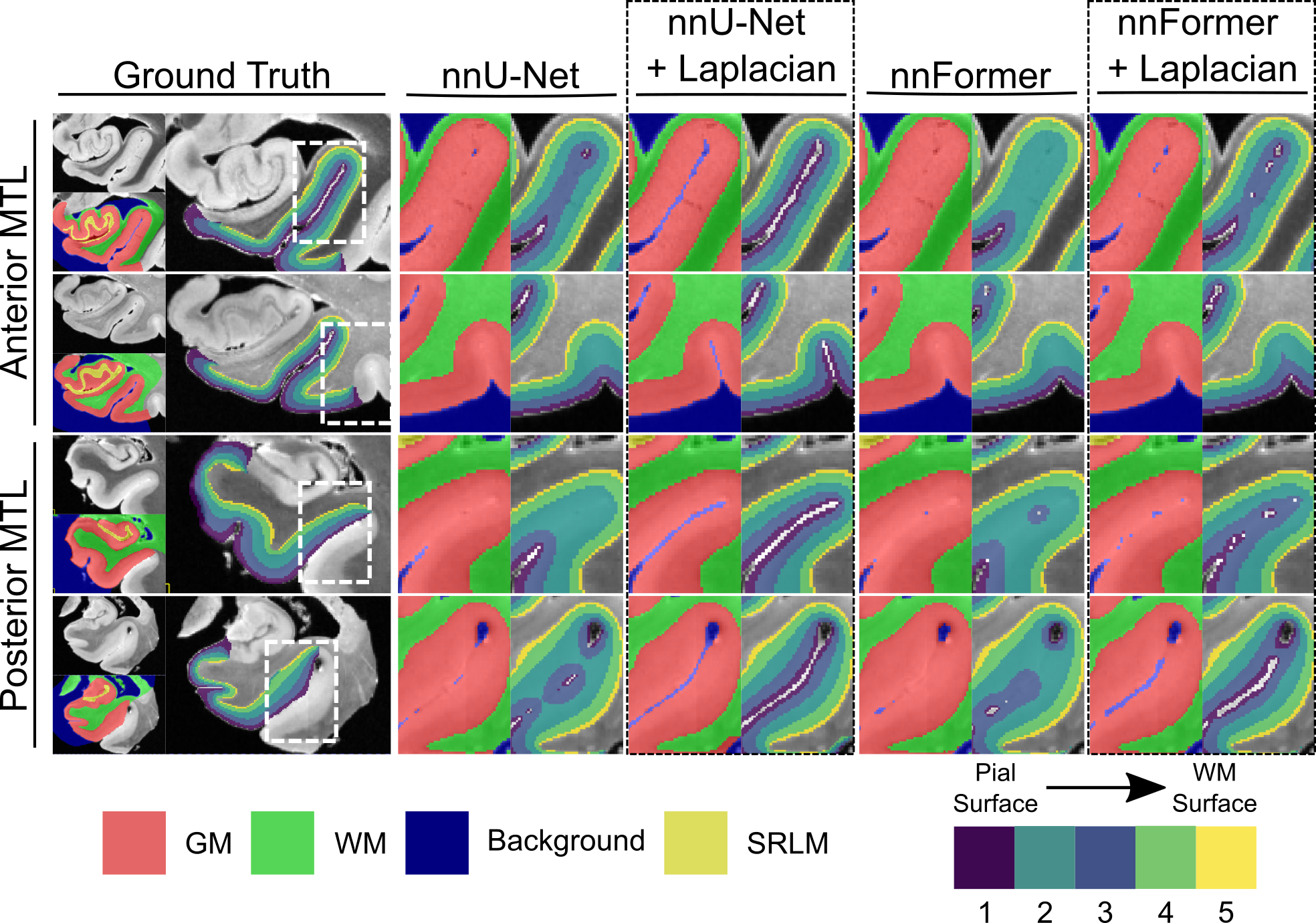}
    \setlength{\belowcaptionskip}{-10pt}
    \setlength{\abovecaptionskip}{-10pt}
    \caption{Qualitative comparison of the cortical segmentations generated by the proposed method and nnU-Net. Cross-sectional views are provided through the anterior and posterior MTL, using four different specimens. The white dashed boxes are used to indicate cortical folds demonstrating improved geometric accuracy. GM: Gray Matter; WM: White Matter; SRLM: Stratum Radiatum Lacunosum Moleculare}
    \label{fig:qual_ressult}
\end{figure}

\begin{table}[ht]
\vspace{-5pt}
\centering
\caption{Quantitative metrics comparing cortical segmentation accuracy of the proposed network and the baseline networks. We compute the Laplacian Segmentation accuracy, which reflects the networks ability to capture the layered nature of the cortex, across the whole MTL, and also separately for the anterior and posterior MTL. Metrics: Dice Score Coefficient (DSC) per label; Hausdorff Distance (HD). Standard deviations are reported in parentheses. (Statistical significance was assessed using paired t-tests; *** < 0.001, **: p < 0.01, *: p < 0.05)}
\label{quant_nnUnet}
\begin{adjustbox}{width={0.9\textwidth}}
\begin{tabular}{|c|c|ccccc|cccc|c|}
\hline
\cellcolor[HTML]{FFFFFF} &  & \multicolumn{5}{c|}{\textbf{DSC Laplacian Segmentation (\%)}} & \multicolumn{4}{c|}{\textbf{DSC Tissue Segmentation (\%)}} &  \\ \cline{3-11}
\multirow{-2}{*}{\cellcolor[HTML]{FFFFFF}\textbf{}} & \multirow{-2}{*}{\textbf{Method}} & \multicolumn{5}{c|}{\textbf{Pial Surface $\rightarrow$ WM Surface}} & \multicolumn{1}{c|}{\textbf{GM}} & \multicolumn{1}{c|}{\textbf{WM}} & \multicolumn{1}{c|}{\textbf{BG}} & \textbf{SRLM} & \multirow{-2}{*}{\textbf{HD 95 (mm)}} \\ \hline
 & nnU-Net & \multicolumn{1}{c|}{\begin{tabular}[c]{@{}c@{}}78.3\\ (3.3)\end{tabular}} & \multicolumn{1}{c|}{\begin{tabular}[c]{@{}c@{}}78.6\\ (3.1)\end{tabular}} & \multicolumn{1}{c|}{\begin{tabular}[c]{@{}c@{}}77.4\\ (3.1)\end{tabular}} & \multicolumn{1}{c|}{\begin{tabular}[c]{@{}c@{}}77.7\\ (5.0)\end{tabular}} & \begin{tabular}[c]{@{}c@{}}58.7\\ (12.3)\end{tabular} & \multicolumn{1}{c|}{\begin{tabular}[c]{@{}c@{}}94.5\\ (1.5)\end{tabular}} & \multicolumn{1}{c|}{\begin{tabular}[c]{@{}c@{}}96.0\\ (1.3)\end{tabular}} & \multicolumn{1}{c|}{\begin{tabular}[c]{@{}c@{}}95.3\\ (6.1)\end{tabular}} & \begin{tabular}[c]{@{}c@{}}85.8\\ (3.8)\end{tabular} & \begin{tabular}[c]{@{}c@{}}0.341\\ (0.134)\end{tabular} \\ \cline{2-12} 
 & \cellcolor[HTML]{C0C0C0} & \multicolumn{1}{c|}{\cellcolor[HTML]{C0C0C0}} & \multicolumn{1}{c|}{\cellcolor[HTML]{C0C0C0}} & \multicolumn{1}{c|}{\cellcolor[HTML]{C0C0C0}} & \multicolumn{1}{c|}{\cellcolor[HTML]{C0C0C0}} & \cellcolor[HTML]{C0C0C0} & \multicolumn{1}{c|}{\cellcolor[HTML]{C0C0C0}} & \multicolumn{1}{c|}{\cellcolor[HTML]{C0C0C0}} & \multicolumn{1}{c|}{\cellcolor[HTML]{C0C0C0}} & \cellcolor[HTML]{C0C0C0} & \cellcolor[HTML]{C0C0C0} \\
 & \multirow{-2}{*}{\cellcolor[HTML]{C0C0C0}nnU-Net + Laplacian} & \multicolumn{1}{c|}{\multirow{-2}{*}{\cellcolor[HTML]{C0C0C0}\begin{tabular}[c]{@{}c@{}}80.7****\\ (2.6)\end{tabular}}} & \multicolumn{1}{c|}{\multirow{-2}{*}{\cellcolor[HTML]{C0C0C0}\begin{tabular}[c]{@{}c@{}}81.8****\\ (2.4)\end{tabular}}} & \multicolumn{1}{c|}{\multirow{-2}{*}{\cellcolor[HTML]{C0C0C0}\begin{tabular}[c]{@{}c@{}}80.4****\\ (3.5)\end{tabular}}} & \multicolumn{1}{c|}{\multirow{-2}{*}{\cellcolor[HTML]{C0C0C0}\begin{tabular}[c]{@{}c@{}}78.9*\\ (5.8)\end{tabular}}} & \multirow{-2}{*}{\cellcolor[HTML]{C0C0C0}\begin{tabular}[c]{@{}c@{}}58.2\\ (13.4)\end{tabular}} & \multicolumn{1}{c|}{\multirow{-2}{*}{\cellcolor[HTML]{C0C0C0}\begin{tabular}[c]{@{}c@{}}94.5\\ (1.7)\end{tabular}}} & \multicolumn{1}{c|}{\multirow{-2}{*}{\cellcolor[HTML]{C0C0C0}\begin{tabular}[c]{@{}c@{}}95.9\\ (1.3)\end{tabular}}} & \multicolumn{1}{c|}{\multirow{-2}{*}{\cellcolor[HTML]{C0C0C0}\begin{tabular}[c]{@{}c@{}}95.4\\ (5.9)\end{tabular}}} & \multirow{-2}{*}{\cellcolor[HTML]{C0C0C0}\begin{tabular}[c]{@{}c@{}}85.6\\ (3.8)\end{tabular}} & \multirow{-2}{*}{\cellcolor[HTML]{C0C0C0}\begin{tabular}[c]{@{}c@{}}0.344\\ (0.205)\end{tabular}} \\ \cline{2-12} 
 &  & \multicolumn{1}{c|}{} & \multicolumn{1}{c|}{} & \multicolumn{1}{c|}{} & \multicolumn{1}{c|}{} &  & \multicolumn{1}{c|}{} & \multicolumn{1}{c|}{} & \multicolumn{1}{c|}{} &  &  \\
 & \multirow{-2}{*}{nnFormer} & \multicolumn{1}{c|}{\multirow{-2}{*}{\begin{tabular}[c]{@{}c@{}}75.5 \\ (4.1)\end{tabular}}} & \multicolumn{1}{c|}{\multirow{-2}{*}{\begin{tabular}[c]{@{}c@{}}75.5 \\ (3.7)\end{tabular}}} & \multicolumn{1}{c|}{\multirow{-2}{*}{\begin{tabular}[c]{@{}c@{}}74.6 \\ (3.4)\end{tabular}}} & \multicolumn{1}{c|}{\multirow{-2}{*}{\begin{tabular}[c]{@{}c@{}}75.9 \\ (4.7)\end{tabular}}} & \multirow{-2}{*}{\begin{tabular}[c]{@{}c@{}}57.0*\\ (10.6)\end{tabular}} & \multicolumn{1}{c|}{\multirow{-2}{*}{\begin{tabular}[c]{@{}c@{}}93.8** \\ (1.9)\end{tabular}}} & \multicolumn{1}{c|}{\multirow{-2}{*}{\begin{tabular}[c]{@{}c@{}}95.1 *\\ (2.1)\end{tabular}}} & \multicolumn{1}{c|}{\multirow{-2}{*}{\begin{tabular}[c]{@{}c@{}}95.1 \\ (6.1)\end{tabular}}} & \multirow{-2}{*}{\begin{tabular}[c]{@{}c@{}}83.8** \\ (4.2)\end{tabular}} & \multirow{-2}{*}{\begin{tabular}[c]{@{}c@{}}0.559 \\ (0.885)\end{tabular}} \\ \cline{2-12} 
\multirow{-6}{*}{\textbf{\begin{tabular}[c]{@{}c@{}}Whole \\ MTL\end{tabular}}} & \cellcolor[HTML]{C0C0C0}nnFormer + Laplacian & \multicolumn{1}{c|}{\cellcolor[HTML]{C0C0C0}\begin{tabular}[c]{@{}c@{}}78.1***\\ (3.4)\end{tabular}} & \multicolumn{1}{c|}{\cellcolor[HTML]{C0C0C0}\begin{tabular}[c]{@{}c@{}}79.8****\\ (3.0)\end{tabular}} & \multicolumn{1}{c|}{\cellcolor[HTML]{C0C0C0}\begin{tabular}[c]{@{}c@{}}78.4****\\ (3.6)\end{tabular}} & \multicolumn{1}{c|}{\cellcolor[HTML]{C0C0C0}\begin{tabular}[c]{@{}c@{}}77.3***\\ (5.2)\end{tabular}} & \cellcolor[HTML]{C0C0C0}\begin{tabular}[c]{@{}c@{}}55.7\\ (11.6)\end{tabular} & \multicolumn{1}{c|}{\cellcolor[HTML]{C0C0C0}\begin{tabular}[c]{@{}c@{}}93.5\\ (2.2)\end{tabular}} & \multicolumn{1}{c|}{\cellcolor[HTML]{C0C0C0}\begin{tabular}[c]{@{}c@{}}94.9\\ (2.3)\end{tabular}} & \multicolumn{1}{c|}{\cellcolor[HTML]{C0C0C0}\begin{tabular}[c]{@{}c@{}}95.1\\ (6.0)\end{tabular}} & \cellcolor[HTML]{C0C0C0}\begin{tabular}[c]{@{}c@{}}83.2\\ (4.5)\end{tabular} & \cellcolor[HTML]{C0C0C0}\begin{tabular}[c]{@{}c@{}}0.589\\ (0.980)\end{tabular} \\ \hline
 &  & \multicolumn{1}{c|}{} & \multicolumn{1}{c|}{} & \multicolumn{1}{c|}{} & \multicolumn{1}{c|}{} &  & \multicolumn{1}{c|}{} & \multicolumn{1}{c|}{} & \multicolumn{1}{c|}{} &  &  \\
 & \multirow{-2}{*}{nnU-Net} & \multicolumn{1}{c|}{\multirow{-2}{*}{\begin{tabular}[c]{@{}c@{}}80.0 \\ (3.7)\end{tabular}}} & \multicolumn{1}{c|}{\multirow{-2}{*}{\begin{tabular}[c]{@{}c@{}}80.6\\ (3.4)\end{tabular}}} & \multicolumn{1}{c|}{\multirow{-2}{*}{\begin{tabular}[c]{@{}c@{}}79.9 \\ (3.3)\end{tabular}}} & \multicolumn{1}{c|}{\multirow{-2}{*}{\begin{tabular}[c]{@{}c@{}}79.6\\ (5.0)\end{tabular}}} & \multirow{-2}{*}{\begin{tabular}[c]{@{}c@{}}59.3\\ (12.8)\end{tabular}} & \multicolumn{1}{c|}{\multirow{-2}{*}{\begin{tabular}[c]{@{}c@{}}94.5\\ (1.6)\end{tabular}}} & \multicolumn{1}{c|}{\multirow{-2}{*}{\begin{tabular}[c]{@{}c@{}}95.0**\\ (1.6)\end{tabular}}} & \multicolumn{1}{c|}{\multirow{-2}{*}{\begin{tabular}[c]{@{}c@{}}95.7\\ (5.1)\end{tabular}}} & \multirow{-2}{*}{\begin{tabular}[c]{@{}c@{}}85.9\\ (4.4)\end{tabular}} & \multirow{-2}{*}{\begin{tabular}[c]{@{}c@{}}0.208\\ (0.025)\end{tabular}} \\ \cline{2-12} 
 & \cellcolor[HTML]{C0C0C0}nnU-Net + Laplacian & \multicolumn{1}{c|}{\cellcolor[HTML]{C0C0C0}\begin{tabular}[c]{@{}c@{}}81.6**\\ (3.0)\end{tabular}} & \multicolumn{1}{c|}{\cellcolor[HTML]{C0C0C0}\begin{tabular}[c]{@{}c@{}}82.8***\\ (2.7)\end{tabular}} & \multicolumn{1}{c|}{\cellcolor[HTML]{C0C0C0}\begin{tabular}[c]{@{}c@{}}81.8***\\ (3.7)\end{tabular}} & \multicolumn{1}{c|}{\cellcolor[HTML]{C0C0C0}\begin{tabular}[c]{@{}c@{}}80.3\\ (5.9)\end{tabular}} & \cellcolor[HTML]{C0C0C0}\begin{tabular}[c]{@{}c@{}}58.1\\ (13.7)\end{tabular} & \multicolumn{1}{c|}{\cellcolor[HTML]{C0C0C0}\begin{tabular}[c]{@{}c@{}}94.4\\ (1.9)\end{tabular}} & \multicolumn{1}{c|}{\cellcolor[HTML]{C0C0C0}\begin{tabular}[c]{@{}c@{}}94.7\\ (1.7)\end{tabular}} & \multicolumn{1}{c|}{\cellcolor[HTML]{C0C0C0}\begin{tabular}[c]{@{}c@{}}95.9\\ (5.0)\end{tabular}} & \cellcolor[HTML]{C0C0C0}\begin{tabular}[c]{@{}c@{}}85.7\\ (4.5)\end{tabular} & \cellcolor[HTML]{C0C0C0}\begin{tabular}[c]{@{}c@{}}0.217\\ (0.049)\end{tabular} \\ \cline{2-12} 
 & nnFormer & \multicolumn{1}{c|}{\begin{tabular}[c]{@{}c@{}}76.8\\ (4.5)\end{tabular}} & \multicolumn{1}{c|}{\begin{tabular}[c]{@{}c@{}}76.9\\ (4.3)\end{tabular}} & \multicolumn{1}{c|}{\begin{tabular}[c]{@{}c@{}}76.5\\ (4.4)\end{tabular}} & \multicolumn{1}{c|}{\begin{tabular}[c]{@{}c@{}}78.0\\ (5.1)\end{tabular}} & \begin{tabular}[c]{@{}c@{}}58.4**\\ (11.4)\end{tabular} & \multicolumn{1}{c|}{\begin{tabular}[c]{@{}c@{}}93.8*\\ (2.5)\end{tabular}} & \multicolumn{1}{c|}{\begin{tabular}[c]{@{}c@{}}93.9\\ (3.6)\end{tabular}} & \multicolumn{1}{c|}{\begin{tabular}[c]{@{}c@{}}95.7\\ (5.2)\end{tabular}} & \begin{tabular}[c]{@{}c@{}}83.7**\\ (5.0)\end{tabular} & \begin{tabular}[c]{@{}c@{}}0.208\\ (0.0205)\end{tabular} \\ \cline{2-12} 
\multirow{-5}{*}{\textbf{\begin{tabular}[c]{@{}c@{}}Anterior \\ MTL\end{tabular}}} & \cellcolor[HTML]{C0C0C0}nnFormer + Laplacian & \multicolumn{1}{c|}{\cellcolor[HTML]{C0C0C0}\begin{tabular}[c]{@{}c@{}}79.3***\\ (3.4)\end{tabular}} & \multicolumn{1}{c|}{\cellcolor[HTML]{C0C0C0}\begin{tabular}[c]{@{}c@{}}80.9****\\ (3.0)\end{tabular}} & \multicolumn{1}{c|}{\cellcolor[HTML]{C0C0C0}\begin{tabular}[c]{@{}c@{}}80.0****\\ (3.8)\end{tabular}} & \multicolumn{1}{c|}{\cellcolor[HTML]{C0C0C0}\begin{tabular}[c]{@{}c@{}}79.1**\\ (5.2)\end{tabular}} & \cellcolor[HTML]{C0C0C0}\begin{tabular}[c]{@{}c@{}}56.8\\ (12.1)\end{tabular} & \multicolumn{1}{c|}{\cellcolor[HTML]{C0C0C0}\begin{tabular}[c]{@{}c@{}}93.6\\ (2.7)\end{tabular}} & \multicolumn{1}{c|}{\cellcolor[HTML]{C0C0C0}\begin{tabular}[c]{@{}c@{}}93.7\\ (3.9)\end{tabular}} & \multicolumn{1}{c|}{\cellcolor[HTML]{C0C0C0}\begin{tabular}[c]{@{}c@{}}95.8\\ (5.1)\end{tabular}} & \cellcolor[HTML]{C0C0C0}\begin{tabular}[c]{@{}c@{}}83.1\\ (5.4)\end{tabular} & \cellcolor[HTML]{C0C0C0}\begin{tabular}[c]{@{}c@{}}0.212\\ (0.030)\end{tabular} \\ \hline
 & nnU-Net & \multicolumn{1}{c|}{\begin{tabular}[c]{@{}c@{}}74.8\\ (6.2)\end{tabular}} & \multicolumn{1}{c|}{\begin{tabular}[c]{@{}c@{}}74.7\\ (5.8)\end{tabular}} & \multicolumn{1}{c|}{\begin{tabular}[c]{@{}c@{}}73.2\\ (5.4)\end{tabular}} & \multicolumn{1}{c|}{\begin{tabular}[c]{@{}c@{}}74.6\\ (6.1)\end{tabular}} & \begin{tabular}[c]{@{}c@{}}57.2\\ (12.9)\end{tabular} & \multicolumn{1}{c|}{\begin{tabular}[c]{@{}c@{}}94.7 \\ (1.6)\end{tabular}} & \multicolumn{1}{c|}{\begin{tabular}[c]{@{}c@{}}96.9\\ (1.4)\end{tabular}} & \multicolumn{1}{c|}{\begin{tabular}[c]{@{}c@{}}94.3*\\ (9.0)\end{tabular}} & \begin{tabular}[c]{@{}c@{}}85.3\\ (3.4)\end{tabular} & \begin{tabular}[c]{@{}c@{}}0.228 \\ (0.096)\end{tabular} \\ \cline{2-12} 
 & \cellcolor[HTML]{C0C0C0}nnU-Net + Laplacian & \multicolumn{1}{c|}{\cellcolor[HTML]{C0C0C0}\begin{tabular}[c]{@{}c@{}}78.9****\\ (4.7)\end{tabular}} & \multicolumn{1}{c|}{\cellcolor[HTML]{C0C0C0}\begin{tabular}[c]{@{}c@{}}79.8****\\ (3.9)\end{tabular}} & \multicolumn{1}{c|}{\cellcolor[HTML]{C0C0C0}\begin{tabular}[c]{@{}c@{}}78.1****\\ (4.1)\end{tabular}} & \multicolumn{1}{c|}{\cellcolor[HTML]{C0C0C0}\begin{tabular}[c]{@{}c@{}}76.8***\\ (6.5)\end{tabular}} & \cellcolor[HTML]{C0C0C0}\begin{tabular}[c]{@{}c@{}}58.0\\ (13.5)\end{tabular} & \multicolumn{1}{c|}{\cellcolor[HTML]{C0C0C0}\begin{tabular}[c]{@{}c@{}}94.6\\ (1.7)\end{tabular}} & \multicolumn{1}{c|}{\cellcolor[HTML]{C0C0C0}\begin{tabular}[c]{@{}c@{}}97.0\\ (1.1)\end{tabular}} & \multicolumn{1}{c|}{\cellcolor[HTML]{C0C0C0}\begin{tabular}[c]{@{}c@{}}94.1\\ (9.0)\end{tabular}} & \cellcolor[HTML]{C0C0C0}\begin{tabular}[c]{@{}c@{}}85.2\\ (3.4)\end{tabular} & \cellcolor[HTML]{C0C0C0}\begin{tabular}[c]{@{}c@{}}0.223\\ (0.062)\end{tabular} \\ \cline{2-12} 
 & nnFormer & \multicolumn{1}{c|}{\begin{tabular}[c]{@{}c@{}}72.7\\ (8.1)\end{tabular}} & \multicolumn{1}{c|}{\begin{tabular}[c]{@{}c@{}}72.8\\ (7.0)\end{tabular}} & \multicolumn{1}{c|}{\begin{tabular}[c]{@{}c@{}}71.1\\ (6.6)\end{tabular}} & \multicolumn{1}{c|}{\begin{tabular}[c]{@{}c@{}}72.7\\ (6.9)\end{tabular}} & \begin{tabular}[c]{@{}c@{}}55.0\\ (11.9)\end{tabular} & \multicolumn{1}{c|}{\begin{tabular}[c]{@{}c@{}}93.9***\\ (2.5)\end{tabular}} & \multicolumn{1}{c|}{\begin{tabular}[c]{@{}c@{}}96.2*\\ (2.0)\end{tabular}} & \multicolumn{1}{c|}{\begin{tabular}[c]{@{}c@{}}93.8*\\ (8.9)\end{tabular}} & \begin{tabular}[c]{@{}c@{}}84.3*\\ (4.0)\end{tabular} & \begin{tabular}[c]{@{}c@{}}0.348\\ (0.430)\end{tabular} \\ \cline{2-12} 
\multirow{-4}{*}{\textbf{\begin{tabular}[c]{@{}c@{}}Posterior \\ MTL\end{tabular}}} & \cellcolor[HTML]{C0C0C0}nnFormer + Laplacian & \multicolumn{1}{c|}{\cellcolor[HTML]{C0C0C0}\begin{tabular}[c]{@{}c@{}}75.4***\\ (8.3)\end{tabular}} & \multicolumn{1}{c|}{\cellcolor[HTML]{C0C0C0}\begin{tabular}[c]{@{}c@{}}77.5****\\ (6.9)\end{tabular}} & \multicolumn{1}{c|}{\cellcolor[HTML]{C0C0C0}\begin{tabular}[c]{@{}c@{}}75.4****\\ (6.6)\end{tabular}} & \multicolumn{1}{c|}{\cellcolor[HTML]{C0C0C0}\begin{tabular}[c]{@{}c@{}}74.3**\\ (7.9)\end{tabular}} & \cellcolor[HTML]{C0C0C0}\begin{tabular}[c]{@{}c@{}}54.0\\ (13.1)\end{tabular} & \multicolumn{1}{c|}{\cellcolor[HTML]{C0C0C0}\begin{tabular}[c]{@{}c@{}}93.2\\ (3.2)\end{tabular}} & \multicolumn{1}{c|}{\cellcolor[HTML]{C0C0C0}\begin{tabular}[c]{@{}c@{}}96.0\\ (2.2)\end{tabular}} & \multicolumn{1}{c|}{\cellcolor[HTML]{C0C0C0}\begin{tabular}[c]{@{}c@{}}93.6\\ (9.0)\end{tabular}} & \cellcolor[HTML]{C0C0C0}\begin{tabular}[c]{@{}c@{}}83.6\\ (4.0)\end{tabular} & \cellcolor[HTML]{C0C0C0}\begin{tabular}[c]{@{}c@{}}0.436\\ (0.692)\end{tabular} \\ \hline
\end{tabular}
\end{adjustbox}
\vspace{-10pt}
\end{table}

\section{Results and Discussion}

\subsection{Segmentation Accuracy}
Table \ref{quant_nnUnet} presents the quantitative results, averaged across four cross-validation folds, evaluating the proposed framework using the corresponding backbone networks as baseline. Since sulci can be very thin structures, any improvements to mislabeled sulci would contribute minimally to the tissue segmentation DSC. Therefore, it is not surprising that we do not see statistically significant differences in the GM DSC measures. However, when looking at the Laplacian segmentation DSC, the results show that the Laplacian solver significantly improves upon the baseline performance of both nnU-Net and nnFormer in terms of better preserving the layered structure of the cortex. This is visualized in Fig. \ref{fig:qual_ressult} which provides a qualitative comparison of the predicted segmentations generated by each method. Labels 1-3 correspond to layers closest to the pial surface, and are therefore the class labels mostly likely to reflect errors such as bridged or unresolved sulci. 

We observe that in the anterior portion of the MTL, the baseline models are often able to distinguish the sulcus, even without the Laplacian constraint (Fig. \ref{fig:qual_ressult}, row 2). This is likely because in the ground truth segmentation protocol, the labeled GM extends to include both banks of the CS in the anterior MTL, but only the medial bank of the CS in the posterior MTL. As a result, in the ground truth segmentation, the sulcus is clearly labeled in the anterior MTL and therefore included in the tissue segmentation loss. Conversely, in the posterior MTL, the ground truth tissue segmentation does not explicitly enforce the presence of the sulcus. However, in this region, the Laplacian segmentation term implicitly includes information about the location of the sulcus and therefore drives the network to learn the correct pial boundary of the cortex. To further investigate the contribution of the proposed loss function in the anterior and posterior MTL, we computed the DSC metrics of the Laplacian segmentations separately for the anterior and posterior MTL (Table \ref{quant_nnUnet}). We observe that even when considering the anterior MTL on its own, the proposed framework improves Laplacian segmentation accuracy compared to the baseline networks, confirming that the addition of the Laplacian term is in fact contributing towards the network better learning the layered organization of the cortex. This is further seen in the \textit{nnU-Net+Laplacian} result in Fig. \ref{fig:qual_ressult}, row 2, where the proposed network is able to detect a buried sulcus in a cortical fold not included within the ground truth region of interest. We note that Laplacian segmentation label 5, which corresponds to the innermost cortical surface at the GM/WM boundary, forms a very thin layer and therefore has greater variation in DSC compared to the other labels.

\subsection{Downstream Thickness Measurements}
\begin{figure}[ht]
    \centering
    \includegraphics[width =\textwidth]{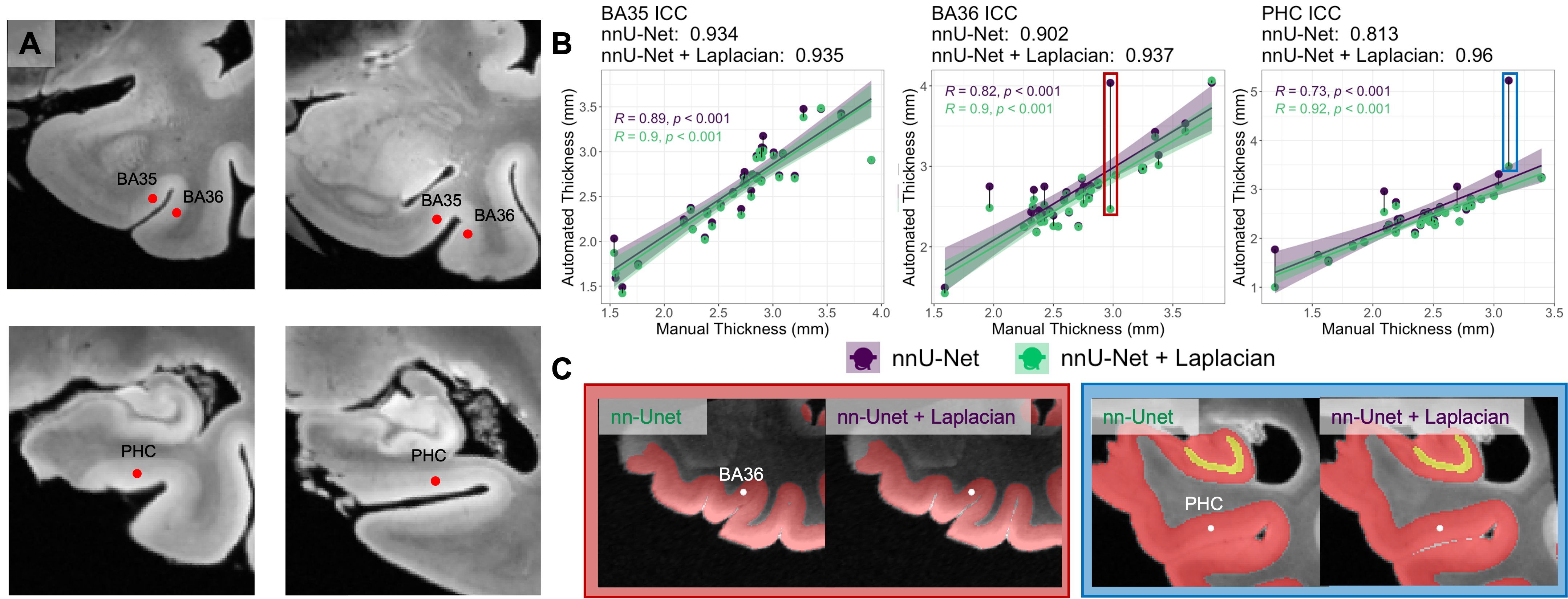}
    \setlength{\belowcaptionskip}{-10pt}
    \setlength{\abovecaptionskip}{-10pt}
    \caption{A) Example scan showing the 6 landmarks where cortical thickness is measured. For each subregion, the thickness measurement is averaged across two landmark locations. B) Scatter plots showing the correlation between automated segmentation-based cortical thickness measurements and reference measurements based on semi-automatic segmentations for three MTL subregions, with (green) and without (purple) the Laplacian constraint. C) Segmentations produced by \textit{nnUnet} and \textit{nnU-Net+Laplacian} for BA36 (red) and PHC (blue) landmarks where thickness measures derived from the two networks differed the most. BA: Brodmann Area; PHC: parahippocampal cortex}
\label{fig:nnUnet_dots}
\end{figure}

Fig. \ref{fig:nnUnet_dots} shows the results of the morphometry analysis correlating automated and manual measurements of cortical thickness in BA35, BA36 and the PHC, when using cortical segmentations generated with and without the Laplacian-based loss term. Since we found that nnU-Net achieves better segmentation performance than nnFormer, we only conducted experiments using nnU-Net in the seconday analysis. Compared to the baseline nnU-Net, we observe that the thickness measurements of BA36 and the PHC computed using the proposed network are more strongly correlated with manual measurements, in terms of both correlation coefficient and ICC. Both models achieve similar correlations in BA35. BA36 is located in the anterior MTL, whereas PHC corresponds to the posterior MTL. The strengthened correlations in both these regions further demonstrate that the proposed method is able to improve the accuracy of the predicted segmentations across the whole length of MTL. 

\section{Conclusions}
\label{conc}
\vspace{-8pt}
We present a novel deep learning-based solution for cortical segmentation, applied to \textit{ex vivo} MRI, that is able to learn the layered geometry of the cortex by locally imposing Laplacian mappings between the predicted WM and pial cortical surfaces. A limitation of this approach is the long run-time of the iterative solver during training ($\sim$9x slower/epoch relative to the backbone). However, at inference time, the input image is only passed through the backbone segmentation network which typically takes 3-5 minutes per scan. Another limitation is the need for the sulci to be well delineated in the training data. In the future, we will explore ways to relax this requirement, perhaps using additional geometric priors. While in this work we demonstrate the utility of our approach in the context of MTL cortical segmentation, this approach can be extended to other high-resolution neuroimaging datasets such as \textit{ex vivo} whole hemisphere scans or other \textit{in vivo} image segmentation tasks which involve similar sheet-like structures. Future work will focus on applying this method to \textit{in vivo} brain MRI, thus allowing for evaluation of our approach against existing cortical surface reconstruction methods.

\vspace{-8pt}
\subsubsection{Acknowledgements} We gratefully acknowledge the tissue donors and their families. This work was supported by the NIH (Grants RF1 AG069474, P30 AG072979 and R01 AG056014), a UCLM travel and research grant (to R.I), and an Alzheimer’s Association grant (AARF-19-615258) (to L.E.M.W).

\squeezeup
%\bibliographystyle{splncs04}

%\bibliography{bibliography}

\end{document}